\documentclass[11pt]{article}
\usepackage{moriond,epsfig,amssymb,graphicx,amsmath,hyperref}

\newcommand{\beq}{\begin{equation}}
\newcommand{\eeq}{\end{equation}}
\newcommand{\rad}{{\cal T}}

\def\beqa{\begin{eqnarray}}
\def\eeqa{\end{eqnarray}}
\begin{document}
\vspace*{4cm}
\title{The radion potential and supersymmetry breaking in detuned RS}

\author{YAEL SHADMI}

\address{Physics Department, Technion--Israel Institute of Technology, 
Haifa 32000, Israel}

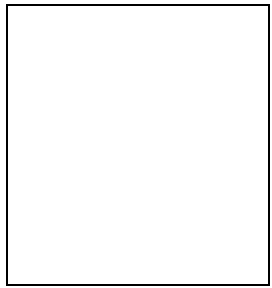
\maketitle\abstracts{We study radiative corrections to the radion potential
in the supersymmetric ``detuned RS model'', with supersymmetry broken 
by boundary conditions. Classically, the radion is stabilized in
this model, and the 4d theory is AdS$_4$. 
With a few bulk hypermultiplets, the one-loop correction to the
cosmological constant is positive. 
For small warping, this correction can (almost) cancel the
classical result. The loop expansion is still
reliable in this limit.
The graviphoton zero-mode, which controls supersymmetry breaking,
is a modulus of the classical theory, but is stabilized
at one-loop. Both unbroken supersymmetry and maximal supersymmetry
breaking are stable ground-states of the quantum theory.
}

\section{Introduction}
An essential ingredient of brane-world models is 
fixing the brane distance, or stabilizing the radion field.
In the ``detuned Randall-Sundrum'' (RS)~\cite{RS} model, 
where the brane tensions are different
from their RS values, this happens automatically~\cite{bagger}. 
The supersymmetric version of this model also allows for supersymmetry
breaking by boundary conditions, unlike in RS~\cite{bagger}. 
Classically, however, the resulting 4d theory has a negative cosmological
constant, so it is not suitable for phenomenology.
Here I will describe quantum corrections to the radion 
potential~\cite{us1,us2}.
Since they require supersymmetry breaking, which is non-local, these
corrections are finite, and in the presence of a few bulk hypermultiplets, give
positive contributions to the potential.
Furthermore, in models with very small warping, the loop corrections
can be very significant, and substantially reduce the classical cosmological
constant.

At low energies, the detuned model gives rise to an effective radion theory
with a superpotential and K\"ahler potential which are very similar to 
those obtained in models of flux compactification~\cite{kklt}, 
but whose origin is purely perturbative.
Of the two scalars in this simple example, only the radion is 
stabilized classically, while its partner is stabilized at one loop.

\section{The classical theory}
The supersymmetric detuned RS model was studied in~\cite{bagger,lalak}.
We will now review its main elements. 
Starting with the RS model and allowing arbitrary brane tensions,
the position of the second brane is determined by the jump conditions,
\beq\label{radius}
R=\frac{1}{2\pi k}
\ln  \frac{(T+T_0)(T+T_\pi )}{(T-T_0)(T-T_\pi )}~,
\eeq
where $T_0$ and $T_\pi$ are the brane tensions, $k$ is the AdS$_5$ curvature,
and $T$ is defined in terms of $k$ and the 5d fundamental scale
as $T\equiv 6 M_5^3 k$.
The theory can be supersymmetrized provided that 
\beq\label{inequality}
\vert T_{0,\pi}\vert \leq T \ .
\eeq
The resulting 4d theory is either AdS$_4$ or Mink$_4$ with the metric
\beq\label{metric}
ds^2=a^2(x_5)\hat g_{\mu \nu}dx^\mu dx^\nu-dx_5^2 \ ,
\eeq
where $\hat g_{\mu \nu}$ denotes the standard AdS$_4$ or Mink$_4$ metric
in Poincare coordinates, and
where the warp factor is given by
\beq
a(x_5)=e^{-k x_5}+\frac{1}{4 k^2 L^2}e^{k x_5}~.
\eeq
Here $L$ is the 4d curvature radius, given by
\beq\label{4dcurv}
\frac{1}{4 k^2 L^2}=\frac{T-T_0}{T+T_0}~.
\eeq
It's easy to see that when the inequality~(\ref{inequality}) is violated,
$L^2$ is negative, so the background is given instead by dS$_4$, 
which is not compatible with supersymmetry.
In fact, the origin of~(\ref{inequality}) is the need to introduce
brane ``mass terms'' for the gravitini in order for the brane plus bulk
action to be supersymmetric.
These brane masses are given by,
\beq\label{tensions}
|\alpha_0|^2 = \frac{T-T_0}{T+T_0} = \frac1{4k^2 L^2}\ ,\ \ \ 
{\rm and}\ \  \ \  
|\alpha_\pi|=|\alpha_0|e^{k\pi R} \ ,
\eeq
so~(\ref{inequality}) must hold. 
Furthermore, 
${\cal N}=1$ supersymmetry is broken for $\arg(\alpha_\pi)\neq\arg(\alpha_0)$.
Clearly, this is not the case in RS, for which $\alpha_{0,\pi}$ vanish.
Working in the ``downstairs'' picture, supersymmetry
breaking translates to breaking by boundary conditions.

Since the 5d theory has a local U(1)$_R$ symmetry under which the
gravitino is charged, a phase difference of $\alpha_0$ and 
$\alpha_\pi$ can be compensated by an $x_5$-dependent U(1)$_R$ transformation.
The fifth component of the graviphoton, which is the U(1)$_R$ gauge field,
is then nonzero. This can also be seen in the low energy effective theory.
At low energies, the theory contains, apart from the 4d SUGRA multiplet,
a chiral ${\cal N}=1$ supermultiplet $\rad$, whose scalar component is
$r+i b$ where $b$ is the zero mode of the fifth component of the graviphoton.
The superpotential and K\"aher potential are
\begin{eqnarray}\label{wk}
W &=& \frac1{1-e^{-2k\pi R}}\,
{M_4^2\over L} \, \left(1-  e^{k\pi R}\,
e^{-3k\pi \rad}\right)\ ,\\
K &=& -3 M_4^2 \ln\left(
\frac{1 - e^{-k \pi (\rad+\bar{\rad})}}{1-e^{-2k\pi R}}
\right) \ ,
\end{eqnarray}
up to ${\cal O}(1/(M_4 L)^2)$ corrections (where $M_4$ is the 4d Planck
scale).
Supersymmetry is then broken if,
\beq\label{superb}
D_\rad W \propto \left(1- e^{i(\phi -3k \pi b)}\right)\,\neq 0\,,
\eeq
namely, for non-zero $b$.
The field $b$, however, is a modulus of the classical theory.
The potential derived from~(\ref{wk}) depends on $r$ only,
with a minimum at $r=R$.

The superpotential of~(\ref{wk}) is of the same form as in KKLT 
models~\cite{kklt}.
Similarly, the K\"ahler potential of~(\ref{wk}) coincides with
the KKLT  K\"ahler potential for small $k$.

\section{Radiative corrections}
\label{radiative}
When supersymmetry is broken, loop corrections can modify the 
4d negative cosmological constant.
Because supersymmetry breaking is non-local, loops are cutoff by
the compactification scale, and are therefore finite.
Since the computation of loop corrections in curved space is 
quite complicated~\cite{norman},
we will consider small detuning, or $1/(kL)<<1$, and perturb
around the RS model. To leading order in $1/L^2$, we will then be able
to use flat space propagators. Supersymmetry will further simplify
the calculation.

\subsection{KK contributions}
\label{KK}
In the supersymmetric RS model, each KK level contains two degenerate
gravitini states, as well as bosonic states from the 5d graviton
and graviphoton. 
Upon detuning the brane tensions, the 4d theory becomes AdS$_4$,
so that the masses of each KK supermultiplet, and in particular
the two gravitini,  are split, with the splitting proportional to $1/L$.
If in addition supersymmetry is broken, by a nonzero phase difference
$\phi$ of the gravitino brane terms $\alpha_0$ and $\alpha_\pi$, the gravitini
masses shift further. For small $\phi$, these shifts are proportional to
$\phi/L$. The factor $1/L$ appears because supersymmetry is
restored as the detuning goes to zero.

For unbroken supersymmetry, or $\phi=0$, the contribution of each
KK supermultiplet to the potential vanishes. 
Therefore, we can calculate the correction to the potential
by considering only the gravitini KK tower,
\beq\label{fermv}
\Delta V(\phi)=
\Delta V_{{\mbox{\tiny{bosons}}}}(\phi)
+  \Delta V_{{\mbox{\tiny{fermions}}}}(\phi)
 =
\Delta V_{{\mbox{\tiny{fermions}}}}(\phi) -
V_{{\mbox{\tiny{fermions}}}}(\phi=0) \ .
\eeq
Writing the $n$-th level gravitini masses as,
\beq\label{masses}
{m^{(n)}}^{\pm}= m^{(n)}_0 \pm \frac{1}{L} \left[ c^{(n)}_1
+ c^{(n)}_{1,SB} \phi^2\right]
+ \frac{1}{k L^2} \left[ c^{(n)}_2  + c^{(n)}_{2,SB} \phi^2\right]
+ {\cal O}(\phi^4)\ ,
\eeq
where the dimensionless coefficients $c$, which depend on $k$ and $R$,
are calculated in~\cite{us2}, and where   $m^{(n)}_0$ is the RS mass,
the one-loop potential is
\beqa\label{vacen}
\Delta V=4 \frac{1}{L^2} \, \phi^2\, \times
\ \ \ \ \ \ \ \ \ \ \ \ \ \ \ \ \ \ \ \ \ \ \ \ \ 
\ \ \ \ \ \ \ \ \ \ \ \ \ \ \ \ \ \ \ \ \ \ \ \ \ 
\ \ \ \ \ \ \ \ \ \ \ \ \ \ \ \ \ \ \ \ \ 
\\
\sum_{n} \int \frac{d^4p}{(2\pi)^4}\,
\Biggl[\,
2\,\frac{ (m_0^{(n)})^2 \, c_1^{(n)}\, c_{1,SB}^{(n)}}
{\left[p^2+ (m_0^{(n)})^2\right]^2} 
-\,\frac{c_0^{(n)}\,  c_{2,SB}^{(n)}}{p^2+(m_0^{(n)})^2}
-\,\frac{ c_1^{(n)}\, c_{1,SB}^{(n)}}{p^2+(m_0^{(n)})^2  }\, \Biggr]~,
\nonumber\eeqa
up to $1/L^4$ terms.
In general this can only be calculated numerically. 
But it is easy to see from~(\ref{vacen}) that the correction is linear
in the supersymmetry breaking scale.
Since supersymmetry breaking is non-local, the result scales as the warp factor
$\exp(-k\pi R)$. It is therefore significant only for small
warping, $kR<<1$. In this limit, the one-loop potential~(\ref{vacen})
can be calculated analytically and reduces to,
\beq\label{kkcorr}
\Delta V = - {3 \zeta(3)\over
2^5\pi^2}
{1\over (\pi R)^4}\, {1\over ( k L)^2} \,\phi^2\ .
\eeq

\subsection{4d radion theory}
We can also calculate the one-loop correction to the potential
using the 4d radion effective theory. The superpotential of~(\ref{wk})
is not renormalized at one-loop. Since it's proportional to the AdS$_4$ 
curvature $1/L$, its contribution to the potential starts as $1/L^2$. 
Therefore, all we need is the one-loop K\"ahler potential
to zeroth order in $1/L$,
that is, the one-loop correction to the RS K\"ahler potential~\cite{rat,falk}.
This correction depends on the radion superfield only through the combination
$\rad+\bar\rad$ and when combined with the superpotential of~(\ref{wk}),
generates a $b$-dependent potential. The $b$ dependence
is of the form $\pm\sin^2(3k\pi b/2)$, 
where the sign depends on the matter content of the theory.
Pure supergravity gives a negative contribution, while hypermultiplets
give a positive contribution.
The simplest hypermultiplet to consider has bulk mass parameter $c=1/2$,
which gives minus a half of the gravity contribution. For four or more
such multiplets, the net contribution is positive.

The loop correction stabilizes the modulus $b$.
The point $b=0$ is then the global minimum of the potential, with
unbroken supersymmetry.
At $b=2/(3k)$, supersymmetry is maximally broken. This is a maximum
along the $b$ direction. 
Nonetheless, it is stable because the $b$ mass is above the BF bound~\cite{bf}
as long as the net cosmological constant is negative.

As we saw above, the loop correction is suppressed for large warping,
because it involves the warp factor $\exp(-k\pi R)$.
For small warping however, it can be significant.
It is easy to see this by thinking about what happens as $k$ is reduced,
for constant $kL$. 
The tree level potential decreases with $k$, because there is no
classical potential in the flat limit. 
The one-loop contribution on the other hand does not go to zero
in this limit, because there is a non-vanishing Casimir energy
even in the flat limit.
In fact, the result~(\ref{kkcorr}) reproduces the Casimir energy of flat
orbifold models with supersymmetry broken by brane superpotentials proportional
to $1/(kL)$ (for small supersymmetry breaking)~\cite{bz}.
Indeed, 
\beq
\frac{V_{\rm loop}}{V_{\rm tree}} \propto \frac{1}{(kR)^2}\ ,
\eeq 
which can easily compete with the loop suppression.
Perturbation theory is still reliable, because higher loop corrections
are suppressed compared to the one-loop result by the usual loop factors,
as long as the number of bulk hypermultiplets is small.

We are thus led to consider the regime 
\beq
\frac{1}{L}<<k<<\frac{1}{R} \ . 
\eeq
In principle, we can make the 4d cosmological constant arbitrarily
small by a suitable choice of $kR$. 
As long as the 4d cosmological constant is negative,
the saddle point with $b=2/(3k)$ is still stable.
For the purposes of model building however, it is sufficient
to reduce the net cosmological constant to below the typical
MSSM contribution, which is roughly a TeV$^4$.
If we embed the MSSM into the model, the soft masses generated
are at most $1/L$. So the relevant question to ask is whether
we can tune the parameters such that the net cosmological constant
is comparable to $1/L^4$. 
Note that the scale $L$ only appears as a common overall factor,
so the only relevant free parameter, when only $c=1/2$ hypermultiplets
are present, is $kR$. In this simple model, the net cosmological cosmological
cannot be made smaller than $1/L^4$~\footnote{In principle, 
$M_4$ and $R$ are independent parameters, but taking $R$ large relative 
to the inverse Planck scale does not help typically}.
But more complicated models with different hypermultiplets may be more
successful in this regard, since they would involve several free parameters,
and the resulting potentials would contain several terms with different $r$
dependence.

\section{Conclusions}
\label{sec:conclusions}
The supersymmetric detuned RS models has several attractive features.
Classically, the radion is stabilized and supersymmetry can be broken
spontaneously. 
With broken supersymmetry, loop corrections can reduce the 
classical 4d curvature.
For small warping, these loop corrections can be very large,
so this setup may be used as a starting point for constructing
(practically) flat extra dimension models.

From a theoretical point of view, the model we considered here
is an interesting example of moduli stabilization in AdS compactifications.
The classical modulus of the theory is stabilized by loop corrections.
Without any additional ingredients, radiative corrections can lift
the tree-level vacuum energy towards zero with all moduli stabilized.
The new ground state is a saddle point, corresponding to a minimum along
the $r$ direction and a maximum along the $b$ direction.
The field $b$ remains stable for arbitrarily small vacuum energy, since its 
mass is above the BF bound.

\section*{Acknowledgments}
I thank Andrey Katz, Michele Redi, and Yuri Shirman for collaboration.
I also thank the organizers of the XLIst Rencontres de Moriond for putting
together an interesting and pleasant meeting.

\end{document}